# Oxygen migration during resistance switching and failure of hafnium oxide memristors

Suhas Kumar,[1]* Ziwen Wang,[2]* Xiaopeng Huang,[1]* Niru Kumari,[1] Noraica Davila,[1] John Paul Strachan,[1] David Vine,[3] A. L. David Kilcoyne,[3] Yoshio Nishi[2] and R. Stanley Williams[1]

*Equal contribution

[1]*Hewlett Packard Labs, 1501 Page Mill Rd, Palo Alto, CA 94304, USA*

[2]*Stanford University, Stanford, CA 94305, USA*

[3]*Lawrence Berkeley National Laboratory, Berkeley, CA 94720, USA*

*Address correspondence to: Suhas.Kumar@hpe.com, Stan.Williams@hpe.com*

**ABSTRACT**

*While the recent establishment of the role of thermophoresis/diffusion-driven oxygen migration during resistance switching in metal oxide memristors provided critical insights required for memristor modeling, extended investigations of the role of oxygen migration during ageing and failure remain to be detailed. Such detailing will enable failure-tolerant design, which can lead to enhanced performance of memristor-based next-generation storage-class memory. Here we directly observed lateral oxygen migration using in-situ synchrotron x-ray absorption spectromicroscopy of $HfO_x$ memristors during initial resistance switching, wear over millions of switching cycles, and eventual failure, through which we determined potential physical causes of failure. Using this information, we reengineered devices to mitigate three failure mechanisms, and demonstrated an improvement in endurance of about three orders of magnitude.*

Metal-oxide memristors, or resistive random access memory (RRAM) switches, in particular utilizing $HfO_x$ as the resistive switching material, have seen significant interest recently for nonvolatile memory and computation applications.[1-5] There has been particular interest in understanding the role of migration of oxygen atoms in determining the operation of memristors.[6-11] Similar recent advances in understanding the localized nanoscale physico-chemical changes underlying resistance switching[4,12-15] have opened up fresh interests into studying the effect of atomic movements on extended device operation and the nanoscale material behavior during eventual failure and possible techniques to mitigate such failure.[16-18]

To enable scanning transmission x-ray microscopy (STXM) measurements, each device was built on a 200 nm low-stress $Si_3N_4$ film suspended over 50 μm x 50 μm holes etched through a silicon substrate.[13] We fabricated crosspoint $HfO_x$ devices with an active area of 2 μm x 2 μm (Figure 1a) by depositing a bottom electrode (15 nm Pt), a blanket layer of 6 nm $HfO_2$, followed by the top electrode (10 nm TiN and 15 nm Pt). Typical current-voltage plots of these devices (Figure 1b) exhibited the well-recognized resistance switching behavior, or pinched hysteresis loop, that characterizes a memristor.[19] During operation, high and low non-volatile resistance states (also called OFF and ON, respectively) were repeatedly accessed using bipolar voltage pulses. STXM experiments were performed using resonantly tuned x-ray beams mostly in the O K-edge region, with spectral resolution of ~70 meV and a beam diameter <30 nm.[20] The device was electrically connected inside the chamber of the system to enable in-situ operation and ON/OFF cycling to emulate ageing of the memristor. The x-ray absorption spectrum of the material stack within a device crosspoint before its operation (Figure 2a) revealed oxygen bonds to both Hf and Ti, suggesting oxidation of Ti upon sputter deposition of TiN onto $HfO_2$ and a resulting mixture of Ti and Hf oxides. We used the absorption of the pre O K-edge at 522 eV to monitor total thickness and other structural modulations (especially electrode distortions), the intensity of the 531 eV peak (the lowest conduction band of the stack) as an indicator of the relative conductivity within the crosspoint,[1,21] and the post O K-edge at 570 eV to determine the local oxygen concentration in the film.

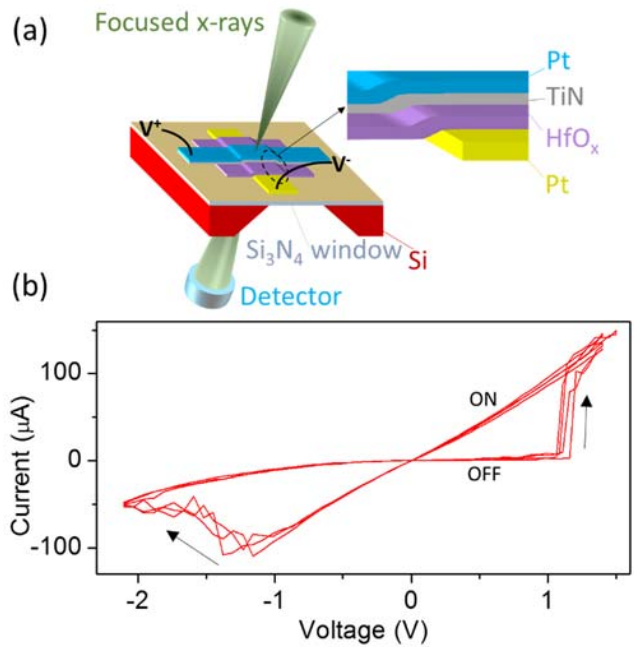

**Figure 1:** (a) Schematic illustration of the crosspoint device and the experimental setup. Inset is an enlargement showing the different materials in the stack. (b) Three successive current-voltage plots of a single device, measured by ramping voltage up and down. The bistable resistance states are marked (ON and OFF).

After the first switching event of a device from its native OFF state to an ON state using a power of ~3 mW (using ~3 V), we imaged the memristor at multiple x-ray energies and clearly observed a ring-like feature within the crosspoint area (Figure 2b). We present an enlarged image of the ring itself in Figure 2d, which shows the logarithmic ratio of images of the ring at 522 eV and 570 eV, which should represent the elemental oxygen concentration without significant contributions from structural variations in the electrodes or other elemental species.[13] In this representation, the ring is evident as an increased oxygen concentration (darker contrast) relative to its surroundings. We then present an image of the ring in Figure 2e, which shows the logarithmic ratio of images of the ring at 570 eV and 531 eV, which is a measure of the density of conduction states normalized to the concentration variation of oxygen.[10,22] This representation reveals that the ring is a region of decreased conductivity (darker contrast). Inspection of Figures 2d and 2e reveals a very small region near the center of the ring with a much lower oxygen concentration and higher electrical conductivity than surrounding areas. To confirm that the ring-like region was indeed the conduction channel, we drove high DC power through the device (~30 mW) until it displayed visible damage (Figure 2c). This damage included severely distorted and missing electrode material, which was likely evaporated because of the localized high temperatures over the conduction channel caused by Joule heating upon application of the DC power, leaving behind a volcano-like topography.

We simulated the radial migration of oxygen during switching using the Strukov model that accounted for coupled thermophoresis and driven-Fick diffusion.[10,23] In this model, thermophoresis drives oxygen radially outward down a Joule-heating-driven temperature gradient from the conductive core, whereas Fick diffusion drives oxygen inward down the resulting concentration gradient, thereby opposing and balancing thermophoresis. We solved the combined equation of continuity for oxygen vacancies $\partial n_V/\partial t = \nabla \cdot J_{Fick} + \nabla \cdot J_{Soret}$, where where $n_V$ is the oxygen vacancy concentration, $J_{Fick}$ and $J_{Soret}$ are fluxes.[10,23] We obtained the a steady-state 2-dimensional oxygen concentration profile (Figure 2f) using a steady-state temperature profile previously derived for similar conducting channels (Figure 2g).[23,24] Figure 2f is in fair agreement with the experimentally measured relative oxygen concentration map in Figure 2d. This exercise of modeling and mapping the conductive channel reasserts the generality of thermophoresis as an important contributor to resistance switching in oxide memristors.

In order to study ageing of HfO$_x$ memristors, we cycled a fresh but identical device over a million times between high and low resistance states using voltage pulses of amplitude 1-2 V (<1 mW), employing an adaptive testing technique described elsewhere (Figure 3a).[13] Eventually, the device failed by getting irreversibly stuck in an ON state after nearly 2 million cycles and was not recoverable even upon application of higher DC electrical powers (up to 30 mW). A scanning electron micrograph (SEM, inset in Figure 3a) of the electrode over the crosspoint area showed extensive damage, confirmed by an atomic force micrograph (AFM, inset in Figure 3a). Further, the x-ray image of the crosspoint at 522 eV displayed significant intensity variations on the length scale of 100 nm (Figure 3b), suggesting widespread morphological changes. Using Ti L-edge spectromicroscopy, we found that the TiN layer was relatively uniform (supplementary material, Figure S5), which indicated that the Pt electrode had likely suffered damage. To investigate the HfO$_x$ films, we mapped the oxygen concentration as previously described, and observed multiple segregated O-rich and O-deficient regions about 100 nm wide (dark and bright areas, respectively, in Figure 3c) scattered across the crosspoint area. For these images, we mapped smaller regions at multiple x-ray energies in order to minimize artifacts due to sample drift that can occur with large-area maps.

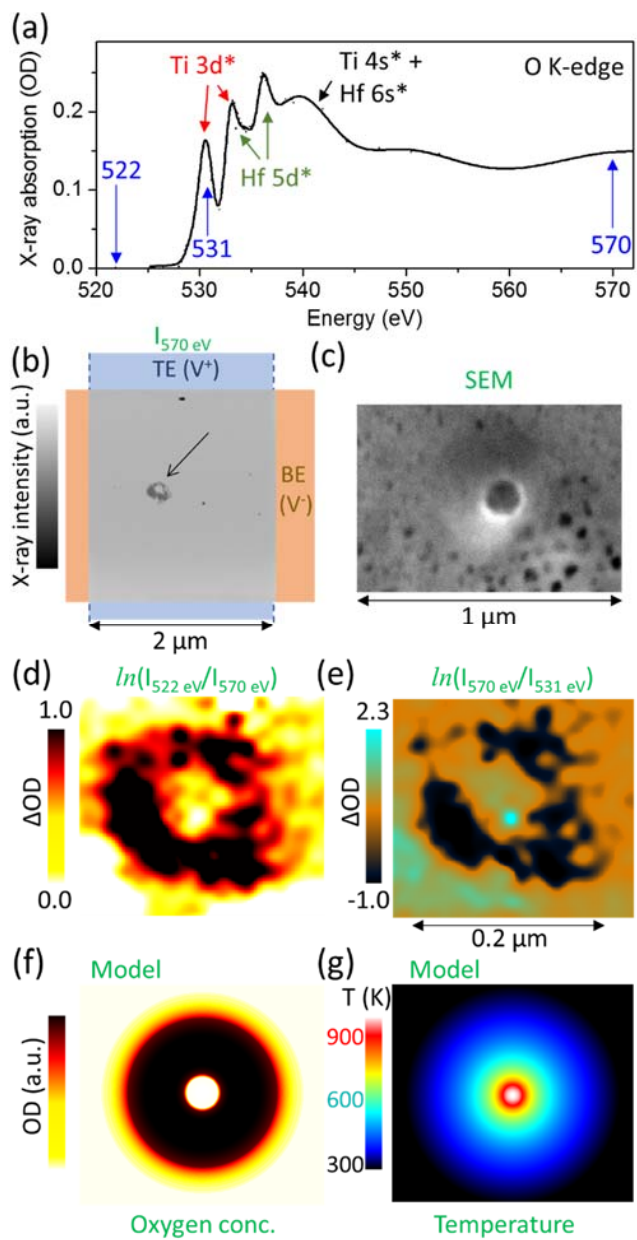

**Figure 2:** (a) O K-edge spectrum of the film within a device before device operation. Energies that are used to quantify different physico-chemical information are marked with blue arrows, while identification of the bands to which the excitations occur are also noted. (b) X-ray image of a device at an energy of 570 eV after one switching event. The black arrow points to a ring-like feature. (c) Scanning electron micrograph (SEM) of the region around the ring of the device in (b) after undergoing failure and damage upon being subjected to high DC powers of ~30 mW. (d) Logarithmic ratio of x-ray images at a pre-absorption-edge energy and at a post-absorption-edge energy, as noted, representing oxygen concentration. (e) Logarithmic ratio of x-ray images at a post-absorption-edge energy and at an on-absorption-edge energy, as noted, representing density of states at the lowest conduction band. Notice that both images (d-e) have a very small region of high brightness inside the ring. (f) Relative oxygen concentration map calculated from the Soret-Fick model and (g) the steady-state temperature profile used for the simulations.

The O K-edge spectra obtained within these bright and dark regions (Figure 3d) confirmed the relative differences in O concentration, and revealed a spectral feature (~534.5 eV) present mainly in the oxygen-rich regions, which appears to be the spectral signature of excess oxygen, possibly as superoxides.[11,25,26] We calculated the most O-rich and O-deficient regions to have oxygen concentrations of +6.5 ± 1% and -6.5 ± 1%, respectively, with respect to the average film composition.[13] Hence, this device exhibited both electrode damage, and lateral movement and segregation of oxygen. The switching failure marked in Figure 3a was likely caused by one or more of the highly O-deficient clusters becoming so conductive that the electrodes became irreversibly shorted.

We then exposed a fresh but identical device to only high DC electrical power of ~30 mW (without cycling between resistance states). This device showed similar electrode damage (Figure 3e), but there was no significant segregation of oxygen (Figure 3f). Millions of switching cycles forced the mobile oxygen atoms to form clusters, which were locally stabilized,[27] whereas, the direct exposure to high DC electrical power did not provide the time and repetition required for cluster formation. In the latter case, the electrode damage followed a single or multiple breakdown events at the edge of the sample, which meant that the $HfO_x$ layer was bypassed.

In order to understand the ageing and failure mechanisms of our devices, we examined ways to improve them. First, in order to provide a more uniform reservoir for excess oxygen and prevent clustering, we introduced a thicker TiN reactive electrode layer. Second, in order to allow the device to sustain large voltage (or power) spikes without as large a temperature rise, we made the Pt electrodes thicker. Third, in order to structurally stabilize the top electrode against delamination, deformation, etc., we introduced a blanket capping layer of $HfO_x+Al_2O_3$ over the top Pt layer. This layer also provided passivation to the active $HfO_x$ layer within the crosspoint area from moisture and related processing chemicals. Thus, the redesigned device stack was: Pt (40 nm)/$HfO_x$ (5 nm)/TiN (50 nm)/Pt (50 nm)/$HfO_x$ (15 nm)/ $Al_2O_3$ (5 nm), with an otherwise identical geometry as before (Figure 4a).

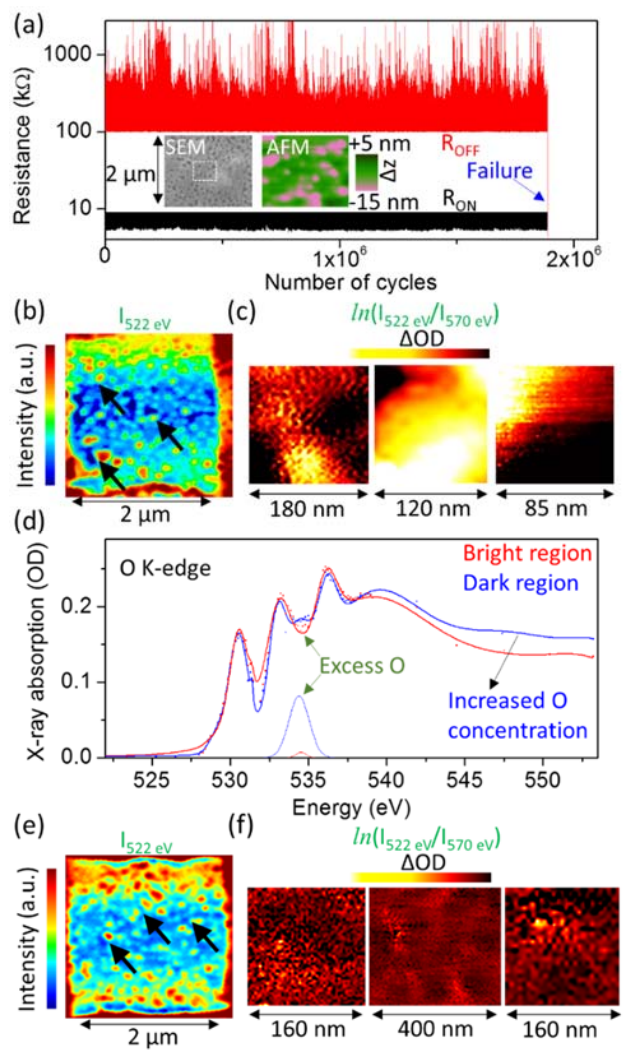

**Figure 3:** (a) ON and OFF resistance states during repeated cycling of a device up to about 2 million times, upon which it fails to switch. Also included are an SEM of the device (inset), after switching failure and attempts to revive the device using high DC power; and a corresponding atomic force micrograph (AFM) (inset) within the dashed rectangle shown in the SEM. The maximum change in topology ($\Delta z$) was 16 nm, which corresponds well to the thickness of the top Pt electrode, implying that most of the damage happened in the top Pt electrode. (b) X-ray map of the device in (a) at 522 eV showing significant non-oxygen contrast. (c) Multiple smaller regions (marked with black arrows in (b)) were chosen to map out the oxygen concentration using logarithmic ratio of maps at different energies, on a common scale, within the region, as noted. (d) O K-edge absorption spectra averaged over two dark and two bright regions (referring to (c)), color coded in the legend. Solid lines are fits from peak decomposition. Only one peak that showed a significant difference is displayed, which possibly corresponds to a superoxide species in a region of high oxygen concentration. (e) X-ray map of another device that underwent failure only due to high DC power (no cycling) at 522 eV, showing significant non-oxygen contrast. (f) Maps of oxygen contrast obtained in a similar way as described above are also shown, displaying much smaller oxygen concentration variations (common data scale with (c)).

Upon cycling five of these devices using the same adaptive switching technique described before, they exhibited repeated switching up to ~$10^9$ cycles (Figure 4b) without any indication of electrical failure, upon which the endurance experiment was terminated. We also applied the same DC power (30 mW) as was done to the thinner device described in Figure 3, but we saw no identifiable damage to the electrodes (inset SEM and AFM of Figure 4b). To illustrate the effect of each of the design modifications listed above, we studied the endurance of different devices as the multiple modification steps noted above were progressively introduced (Figure 4c). Every modification improved the endurance by a measurable quantity, while providing thicker Pt electrodes appeared to introduce the most significant improvement.

Using *in-situ* STXM, we showed that resistive switching, ageing and device failure in metal oxide memristors involve oxygen migration driven by thermophoresis and/or diffusion. We demonstrate mitigation of some of the failure mechanisms by understanding their microphysical origins.

See supplementary material for device fabrication, spectral processing, simulation of oxygen migration due to thermophoresis/diffusion and extended data sets.

All synchrotron measurements were performed at the Advanced Light Source, beamlines 5.3.2.2 and 11.0.2, at Lawrence Berkeley National Laboratory, Berkeley, CA, USA. The Advanced Light Source is supported by the Director, Office of Science, Office of Basic Energy Sciences, of the U.S. Department of Energy under Contract No. DE-AC02-05CH11231. Work was performed in part at the Stanford Nanofabrication Facility which is supported by National Science Foundation through the NNIN under Grant ECS-9731293.

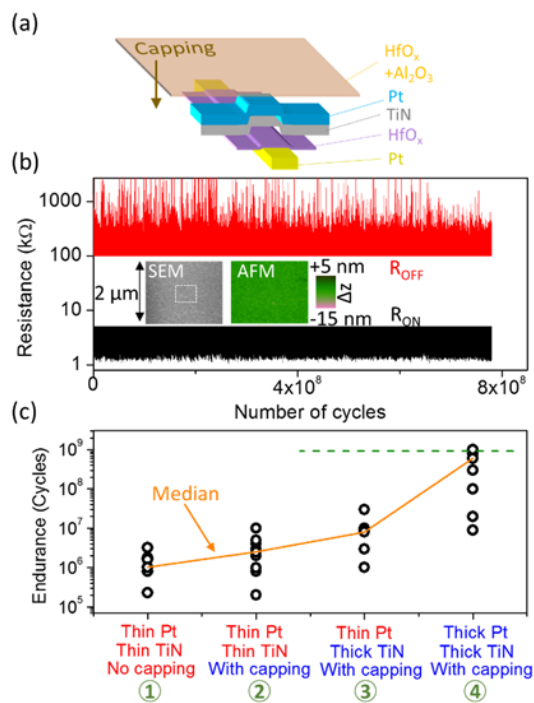

**Figure 4:** (a) Schematic of the redesigned device structures with much thicker TiN and Pt electrodes, and a blanket capping layer on top of the stack. (b) ON and OFF resistance states during repeated cycling of a device from the redesigned batch with a modified material stack which switched ~$10^9$ times and showed no electrical failure. Also included are an SEM of the device (inset) after cycling and subjecting it to high DC power, and a corresponding AFM (inset) within the dashed rectangle shown in the SEM, neither of which show visible damage. (c) Endurance measurements from several sets of devices that were progressively re-designed to incorporate capping, thicker TiN and thicker Pt. The medians of the endurance in each set are connected by a solid orange line. Horizontal green dashed line is the endurance limit of the experiment.